\documentclass[%
 aip,
 amsmath,amssymb,
author-year,%
]{revtex4-1}

\usepackage{graphicx}
\usepackage{dcolumn}
\usepackage{bm}

\usepackage[utf8]{inputenc}
\usepackage[T1]{fontenc}
\usepackage{mathptmx}
\usepackage{etoolbox}
\usepackage{makecell}
\usepackage{subfig}

\makeatletter
\def\@email#1#2{%
 \endgroup
 \patchcmd{\titleblock@produce}
  {\frontmatter@RRAPformat}
  {\frontmatter@RRAPformat{\produce@RRAP{*#1\href{mailto:#2}{#2}}}\frontmatter@RRAPformat}
  {}{}
}%
\makeatother
\begin{document}

\preprint{}

\title[DNS of transition under free-stream turbulence]{Direct numerical simulation of transition under free-stream turbulence and the influence of large integral length scales}

\author{Kristina Đurović}
 \altaffiliation[]{FLOW, Department of Engineering Mechanics, KTH Royal Institute of Technology, SE-100 44 Stockholm, Sweden}
\author{Ardeshir Hanifi}
\email{hanifi@kth.se}
 \altaffiliation[]{FLOW, Department of Engineering Mechanics, KTH Royal Institute of Technology, SE-100 44 Stockholm, Sweden}
\author{Philipp Schlatter}
 \altaffiliation[]{FLOW, Department of Engineering Mechanics, KTH Royal Institute of Technology, SE-100 44 Stockholm, Sweden}
\author{Kenzo Sasaki}
\altaffiliation[]{FLOW, Department of Engineering Mechanics, KTH Royal Institute of Technology, SE-100 44 Stockholm, Sweden}
\author{Dan S. Henningson}
\altaffiliation[]{FLOW, Department of Engineering Mechanics, KTH Royal Institute of Technology, SE-100 44 Stockholm, Sweden}

\date{March 11, 2024}

\begin{abstract}
Under action of free-stream turbulence (FST), elongated streamwise streaky structures are generated inside the boundary layer, and their amplitude and wavelength are crucial for the transition onset. While turbulence intensity is strongly correlated with the transitional Reynolds number, characteristic length scales of the FST are often considered to have a slight impact on the transition location. However, a recent experiment by \cite{fransson_shahinfar_2020} shows significant effects of FST scales. They found that, for higher free-stream turbulence levels and larger integral length scales, an increase in the length scale postpones transition, contrary to established literature. 
Here, we aim at understanding these results by performing a series of high-fidelity simulations. These results provide understanding why the FST integral length scale affects the transition location differently. These integral length scales are so large that the wide streaks introduced in the boundary layer have substantially lower growth in the laminar region upstream of the transition to turbulence, than streaks induced by smaller integral length scales. The energy in the boundary layer subsequently propagate to smaller spanwise scales as a result of the non-linear interaction. When the energy has reached smaller spanwise scales larger amplitude streaks results in regions where the streak growth are larger. It takes longer for the energy from the wider streaks to propagate to the spanwise scales associated with the breakdown to turbulence, than for the those with smaller spanwise scales. Thus there is a faster transition for FST with lower intergral length scales in this case.
\end{abstract}

\maketitle
\section{Introduction}
Laminar-turbulent transition induced by free-stream turbulence is highly relevant due to its occurrence in many practical situations like turbo-machinery flows, mixers, aerodynamics and experiments in conventional wind tunnels. In boundary-layer flows subjected to free-stream turbulence intensities of $1 \%$ or more, transition usually occurs earlier, bypassing the classical scenario triggered by the Tollmien–Schlichting waves. 
This type of flow has been extensively studied numerically in the past \citep{JacobsDurbin, brandt2004, ZakiDurbin, durbin2007transition},
but always using certain simplifications compared to the real setup (e.g. removing the leading edge, simplified FST generation etc.). With the availability of large-scale computers and corresponding simulation codes, we can now perform computations of the complete physical case, extracting flow details in unprecedented detail and putting the focus on details potentially neglected previously. Experimentally these flows have been studied extensively as well, among other by \cite{Matsubara, fransson2004e, fransson2005transition, fransson_shahinfar_2020}. FST-induced transition is characterised by the occurrence of streamwise elongated streaky structures inside the boundary layer. As these streaks travel downstream, they break down into turbulent spots due to their secondary instability. In addition, \cite{Brandt2,Nolan_PIV} found that there may be an interaction between two subsequent
streaks that leads to transition, where the leading edge of a high-speed streak seems to collide with the tail of a low-speed streak. The spots created by the breakdown grow and merge until the flow is fully turbulent.

Despite vast amount of the literature on the subject, the effects of the characteristics of FST on the integral length scale on transition is not clearly understood. Contrary to the much of the established literature, \cite{fransson_shahinfar_2020} showed that the integral length scale of the free-stream turbulence affects the transition location differently depending on its intensity level. An increase in length scale advances transition for low intensities, in agreement with established results. For higher intensities, an increase in length scale postpones transition. In the experimental setup, they observed both trends and made an a semi-imperical model for the trend change. They suggest that one has overlooked the change of the spanwise scale and thus the aspect ratio of the streaks associated with different integral length scales as transition is approached. These results differ with many previous results in that the streaks do not immediately adapt to the boundary layer thickness independent of the free-stream turbulence characteristics. 



A crucial elements for understanding effects of FST on transition is the receptivity at the leading edge which provides a route for coupling vortical disturbances to the boundary layer. Here, the shape of leading edge will also be important. A comparison of transition scenarios in the presence of thin and blunt leading edges has been made by e.g. \cite{nagarajan2007l} showing that the blunt leading edge led to the formation of instability wavepackets, whereas for lower bluntness slowly meandering streaks resulted. In order to be able to study these effects, an accurate computational modeling of the experimental setup needs to be available, including details of the geometry (both plate and wind-tunnel ceiling), boundary conditions and free-stream turbulence. For instance, the numerical FST generation mimics the isotropic, homogeneous free-stream turbulence generated behind the grid in the wind tunnel. Numerous investigations have been presented regarding optimal methods for the generation of unsteady turbulent fields \citep{piomellikeating,TABOR2010553,nitin}. See \cite{annurev-fluid-wu} for a detailed reviews of the techniques used in the literature. 
Here, we study the effects of the FST integral length scale on the transition in an incompressible flat-plate boundary layer using direct numerical simulation (DNS), considering two different cases. The numerical setup corresponds to the experimental investigations by \cite{fransson_shahinfar_2020}. FST is prescribed in the volume section close to the inlet boundaries as a superposition of Fourier modes following \cite{brandt2004} and introduced through a body force. The numerical methodology used allows us to define the energy spectrum of the turbulent inflow with a given integral length scale and turbulence intensity. 

The paper is organized as follows: in \S \ref{sec:flow_case} we describe the employed flow configuration and in \S \ref{sec:num_approach} the numerical methods. Our findings are reported in \S \ref{sec:results} and \S \ref{Influence} and then summarized in \S \ref{sec:conclusion}.
\section{Flow configuration} \label{sec:flow_case}

\begin{figure}
    \centering
    \includegraphics[width=0.8\textwidth]{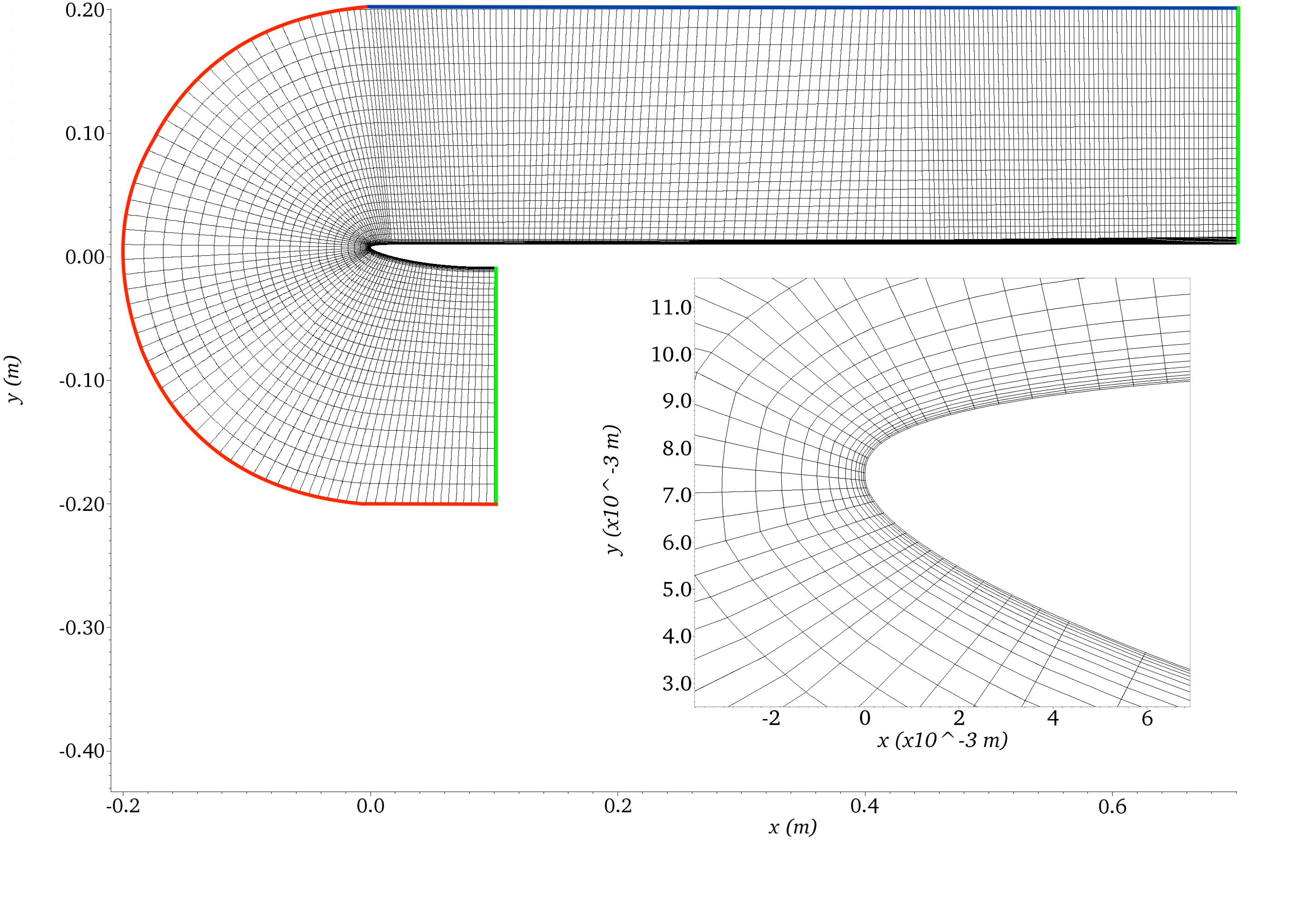}
    \caption{Spectral element mesh used in the present study. Color lines denote different boundary conditions used in numerical simulations, where the red color indicates inlet boundary with imposed Dirichlet condition, the blue line indicates the top boundary with mixed boundary condition from equation (\ref{eq:on}), the green line represents outflow condition as in equation (\ref{eq:outflow}) and the black line represents no-slip boundary condition at the wall. Note that the 2D mesh is replicated in the spanwise direction with size 0.15m.}
    \label{fig:mesh}
\end{figure}


Here, we investigate flow over a flat plate with leading edge where the geometry corresponds to the one used in the experimental work by \cite{fransson_shahinfar_2020}. Free-stream velocity is $U_\infty=6\ m/s$, dynamic viscosity $\mu=1.511\times 10^{-5}\ kg/(ms)$ and density $\rho=1.204\ kg/m^3$. This choice of parameters gives a Reynolds number slightly higher than in the experiments by \cite{fransson_shahinfar_2020}.

The numerical domain is schematically shown in figure \ref{fig:mesh}. The dimensions of the domain in the wall-normal direction range from $y_{min}=-0.2\ m$ to $y_{max}=0.2\ m$, and the plates leading edge is positioned around $x=0\ m,y=0\ m$. The spanwise extension of the domain is from $z_{min}=-0.075\ m$ to $z_{max}=0.075\ m$. Streamwise domain size starts at $x_{min}=-0.2\ m$ and ends at $x_{max}=0.7\ m$. The spectral element mesh is structured and orthogonal near the plate surface. The number of spectral elements is 488320. Their distribution at $z=0\ m$ plane is shown in figure \ref{fig:mesh}. Note that only spectral elements are shown, where the total number of points is obtained by multiplying each direction by the polynomial order ($N=11$) of the simulation. The grid resolution at the plate surface varies with the streamwise location, with the criterion from $\Delta x^+=0.2-6$.  Element distribution increases in the wall-normal direction, where it is the smallest on the plate surface, with $y^+ =0.3 $ and this increases through the boundary layer region and into the free-stream. Resolution in the spanwise direction is $\Delta z^+=4$. Scaling is provided in the viscous units, where $l^{*} =\nu /u_{\tau}$ is viscous length, $\nu$ is the fluid kinematic viscosity and $u_{\tau}=\sqrt{\tau_w/\rho}$, with $\rho$ being the fluid density.  In figure \ref{fig:mesh} enlarged view of the mesh around the sharp leading edge is shown. Parameters used in this study are presented in table \ref{tab:cases}.

\begin{table}
 \caption{Parameters for cases studied here. Total time refers to the snapshots available in the database, where the actual simulation time is longer. $x_{min}=-0.2m$, $x_{max}=0.7m$, $y_{min}=-0.2m$,$y_{max}=0.2m$, $z_{min}=-0.075m$, $z_{max}=0.075m$.}
  \label{tab:cases}

\begin{ruledtabular}
  \begin{tabular}{l c c c c c}
      Case  & \makecell{$T_u$\\ $(\%)$} & \makecell{$\Lambda_x$\\ $(mm)$} & \makecell{Domain size\\ $L_x\times L_y\times L_z$ $(m)$} & \makecell{Resolution\\ $n_x\times n_y\times n_z$}  &   \makecell{Total time \\ $(s)$} \\[3pt]
     \hline
       C1   & $3.40$ & $29.22$ & $0.7\times0.2\times0.15$ & $218\times35\times64$ & 0.4755\\
       C2   & $3.45$ & $11.53$ &  $0.7\times0.2\times0.15$& $218\times35\times64$& 0.6475\\
  \end{tabular}
 \end{ruledtabular}
\end{table}

\section{Numerical approach} \label{sec:num_approach}
\subsection{Equations}
The flow of an incompressible fluid with uniform properties is analysed using Navier-Stokes equations
\begin{equation} \label{eq:NS}
\frac{\partial \mathbf{u}}{\partial t} + \mathbf{u} \cdot \nabla \mathbf{u} = -\nabla p + \frac{1}{Re} \nabla^2 \mathbf{u} + \mathbf{f}, \qquad \nabla \cdot \mathbf{u} = 0, 
\end{equation}
where $\mathbf{u}= \left \{ u,v,w \right \}^T$ represents the velocity vector, $p$ the pressure and $\mathbf{f}$ the body force. The equations (\ref{eq:NS}) are integrated using the Nek5000 \citep{nek5000} code which is based on a spectral-element method \citep{PATERA1984468}. The spatial discretization is done using the Galerkin approximation, following the $\mathbb{P}_N-\mathbb{P}_{N-2}$ formulation. The solution is interpolated within a spectral element employing Lagrange interpolants of orthogonal Legendre polynomials on the Gauss-Lobatto-Legendre quadrature points. The non-linear terms are treated explicitly by third-order extrapolation, whereas the viscous terms are treated implicitly by a third-order backward differentiation scheme. Dealiasing of the non-linear terms is performed using overintegration.

\subsection{Boundary conditions}
As inflow, we use Dirichlet boundary conditions
\[
(u,\ v,\ w)=(u_{ns},\ v_{ns},\ 0).
\]
In the spanwise direction, a periodic condition is enforced, while at the wall, the velocity field is subjected to a no-slip condition. The outflow conditions are the following
\begin{equation} \label{eq:outflow}
 \frac{1}{Re} \frac{\partial u}{\partial x} -p = -p_a, \qquad \frac{\partial v}{\partial x} = 0, \qquad \frac{\partial w}{\partial x} = 0,
\end{equation}
where
\[
p_a=p_{ns}-\frac{1}{Re}\frac{\partial u_{ns}}{\partial x}.
\]
On the top boundary, we employ
\begin{equation} \label{eq:on}
 u = u_{ns} \qquad \frac{1}{Re} \frac{\partial v}{\partial y} -p = -p_a, \qquad w = 0.
\end{equation}

Quantities detonated with subscript $ns$ are obtained through a precursor two-dimensional simulation of the experimental setup with the flat plate, including the trailing flap, placed inside the wind tunnel. The angle of trailing flap was tuned to match the experimental pressure distribution close to the leading edge.

Boundary conditions are marked in figure \ref{fig:mesh}. Red color indicates inlet boundary with imposed Dirichlet condition, blue line indicated the top boundary with mixed boundary condition from equation (\ref{eq:on}), the green line represents outflow condition as in equation (\ref{eq:outflow}) and the black line represents no-slip boundary condition at the wall.
 


\subsection{Free-stream turbulence} 


Here, we use a method similar to that by \cite{Schlatter2001DirectNS, brandt2004} to generate the free-stream turbulence. The free-stream turbulence is prescribed as a superposition of Fourier modes with a random phase shift which can be written as:
 \begin{equation}
    u_i(x,y,z,t)= \sum_{k_x,k_y,k_z} A\hat{u}_i(k_x,k_y,k_z,t)e^{i(k_xx+k_yy+k_zz)}
\end{equation}
where $A$ represents the scaled amplitude of the free-stream modes and $k_x, k_y, k_z$ are streamwise, wall-normal and spanwise wavenumbers, respectively. Using Taylor's frozen turbulence hypothesis, we model the time dependency of perturbations as $e^{i\omega t}$ where $\omega=k_x U_\infty$ is the angular frequency of disturbance. We specify the maximum and minimum amplitudes of the wavenumber vector. The wavenumber space between the minimum and maximum is divided into a set of concentric shells, with each shell representing the amplitude of the three-dimensional wavenumber vectors lying on that shell. The amplitude of the free-stream modes on each spherical shell is scaled using the von K\'arm\'an spectrum given as
\begin{equation}
E(\kappa)=\frac{2}{3}\Lambda\frac{1.606(\kappa \Lambda)^4}{\Big[1.350+(\kappa \Lambda)^2\Big]^{17/6}} q.
\label{eq:vonKarmaneq}
\end{equation}
%
Here, $\Lambda$ represents the nominal integral length scale, which is user specified, $q$ is the total turbulent kinetic energy. 
%

The numerical methodology implemented allows us to define the energy spectrum of the turbulent inflow so that it is possible to investigate the effect of the integral length scale and turbulence intensity of the free-stream turbulence on boundary layer transition. Note that the integral length scale in this equation represents a nominal value and the actual achieved value in the flow has to be checked in the flow itself. 
A detailed description of the method and effects of different numerical parameters is given in \cite{Durovic_FST_generation}.
In the current work, we have used 80 shells with 20 wavenumbers in each shell with $|\mathbf{k}|\in[43.77, 1660]\ m^{-1}$. The minimum spanwise wavenumber ($\beta_{0}=2\pi/0.15\ m^{-1} \approx 42\ m^{-1}$) corresponds to the spanwise extension of the numerical domain and.

The free-stream turbulence discussed above is applied by adding a forcing term on the right-hand side of the momentum equations in the following form

\begin{equation} \label{eq:fringe}
\mathbf{f}(x,y,z,t) = \lambda \left [ \mathbf{U}_{b}(x,y,z) + \mathbf{u}_{dist}(x,y,z,t) - \mathbf{u}(x,y,z,t) \right ].
\end{equation}

Here, $\mathbf{U}_b$ is the baseflow velocity, $\mathbf{u}_{dist}$ are generated disturbances, $\mathbf{u}$ is the local velocity, and $\lambda$ is a fringe function. The fringe function has the following form:

\begin{equation}
    \lambda (x)=\lambda_{max} \left [ S \left ( \frac{x-x_{start}}{\Delta_{rise}} \right ) -S \left ( \frac{x-x_{end}}{\Delta_{fall}} + 1 \right ) \right ]
\end{equation}
where $\lambda_{max}$ represents the maximum strength of the forcing function, $x_{start}$ to $x_{end}$ the spatial extent of the region where the forcing function is applied, $\Delta_{rise}$ and $\Delta_{fall}$ the rise and fall distance of the forcing function and S is a smooth step function defined as
\begin{equation}
S(\xi) = \left\{
\begin{array}{ll}
0\,,                                                 & \quad\xi\le 0\,,\\
\left[1+\exp{\left(\displaystyle{\frac{1}{\xi-1} + \frac{1}{\xi}}\right)}\right]^{-1}, & \quad 0 < \xi < 1\,, \\
1\,,                                                 & \quad \xi\ge 1\,.
\end{array}
\right.
\label{eq:step-function}
\end{equation}
The same shape of forcing function has been used in \cite{Schlatter2001DirectNS}. Values used in the present numerical study are following: $\lambda_{max}=1350$, $x_{start}=-0.185$, $x_{end}=-0.165$, $\Delta_{rise}=0.002$ and $\Delta_{fall}=0.002$. Note that stated parameter values are chosen in the way that gives imposed FST parameters ($Tu_F$ and $\Lambda$) closer to the values of calculated $Tu$ and $\Lambda_{x}$. 


\section{Results}\label{sec:results}

\subsection{Free-stream turbulence}

\begin{figure}
    \centering
\subfloat[]{\includegraphics[trim={0cm 0cm 1cm 1cm},clip,width=0.48\textwidth]{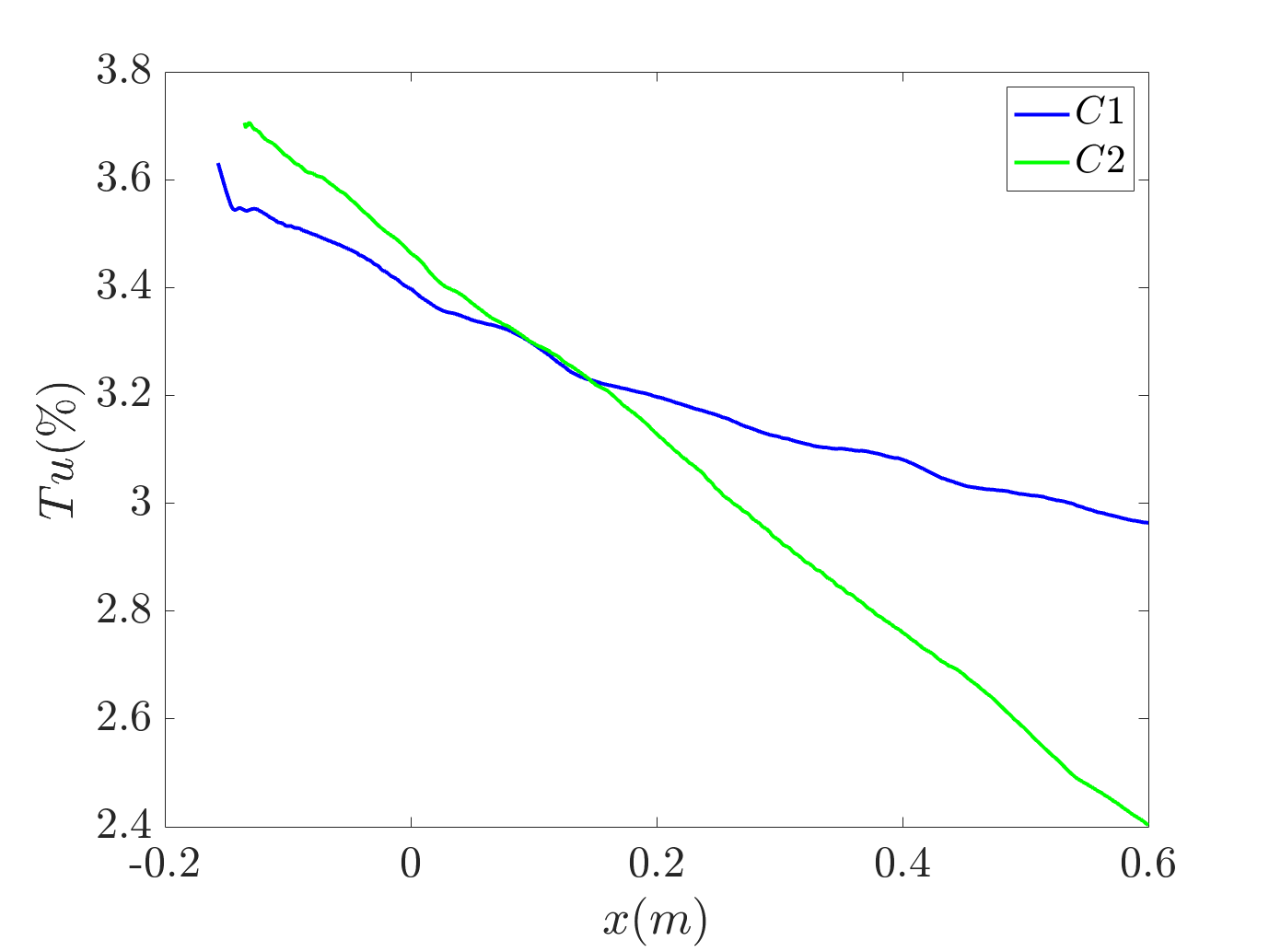}}
\subfloat[]{\includegraphics[trim={0cm 0cm 1cm 1cm},clip,width=0.48\textwidth]{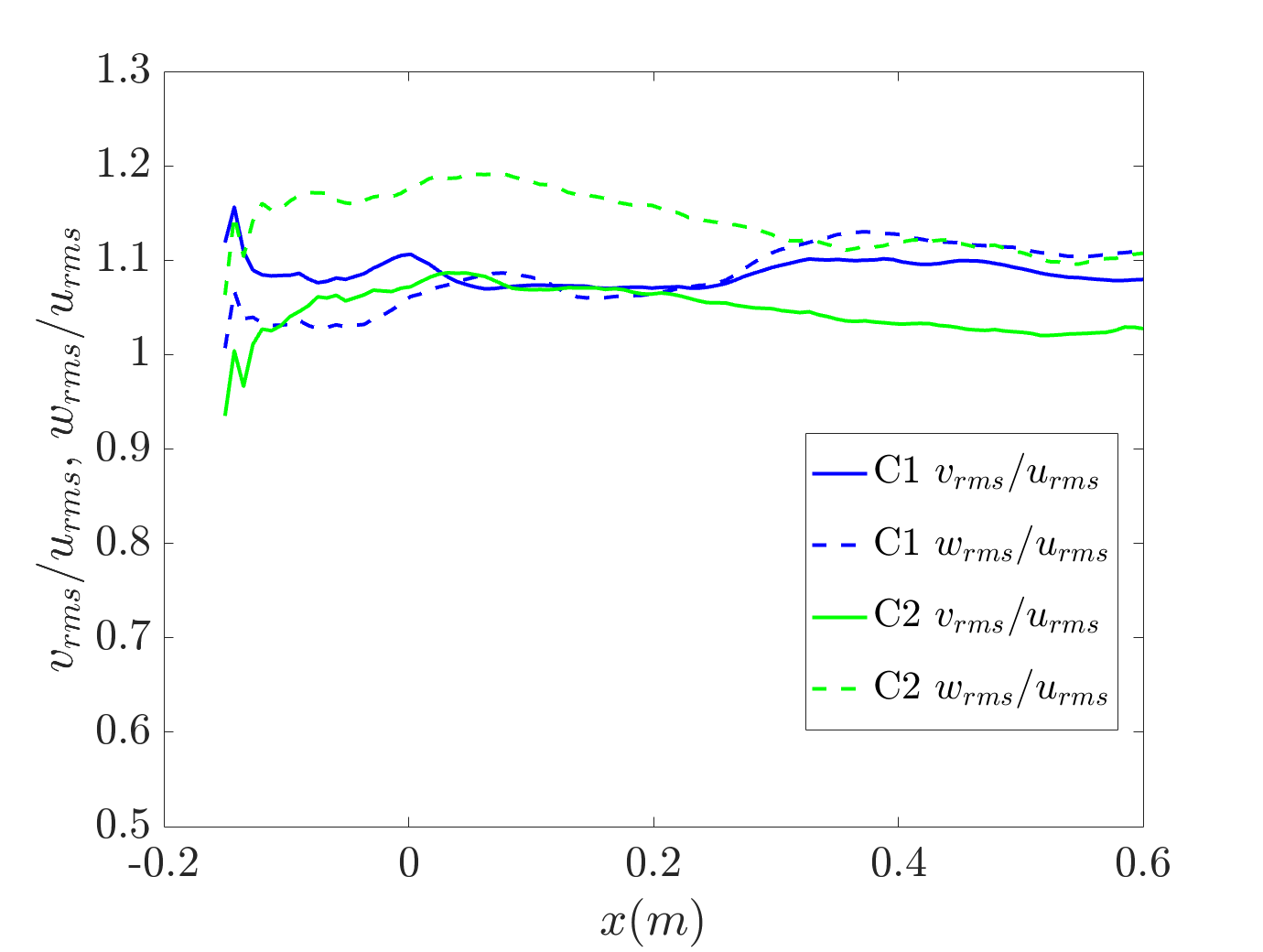}}
   \caption{(a) Turbulence intensity decay in the streamwise direction for different flow cases. (b) Measure for the isotropy of the imposed free-stream turbulence is given in relation to the different velocity fluctuations $v_{rms}/u_{rms}$ and $w_{rms}/u_{rms}$.}
    \label{fig:turms}
\end{figure}

Here, we characterize the generated turbulence field in the free stream. We start by examining behavior of the turbulence intensity, $Tu$, defined as
\begin{equation}
    Tu=\frac{\sqrt{\frac{1}{3}  \left ( \overline{{u'}^2} +\overline{{v'}^2} +\overline{{w'}^2} \right ) }}{U_{\infty}},
\end{equation}
where $u_i'=u_i-\Bar{u}_i$ with $\Bar{u}_i$ being time and spanwise averaged velocities. The evolution of the turbulence intensity in the streamwise direction is shown in figure \ref{fig:turms}a for two cases. Isotropic grid turbulence is known to follow a decay as a  power law where the decay rate is between 0.5 and 1 \citep{Westin}. By performing a least-square fit  of type
\begin{equation}
Tu=C(x-x_0)^{-b},
 \end{equation}
we obtain a decay rate, $b$, between 0.8 and 0.85, depending on the case. This confirms that the decay of the free-stream turbulence is similar to the grid turbulence decay rate. Forced turbulence intensity in the volume forcing region is different for every case, where the goal was to have the same turbulence intensity value at the streamwise position of the leading edge. At that position, both curves have a close values but continue their decay with different decay rates due to different integral length scales. It can be noted that the case with the highest integral length scale (C1) has the slowest decay rate.\medskip

We measure the isotropy of the generated free-stream turbulence by comparing different components of the velocity fluctuations. Figure \ref{fig:turms}b shows the evolution of the averaged fluctuation intensity $u_{rms}$, $v_{rms}$ and $w_{rms}$, defined as
\begin{equation}
    u_{rms}=\sqrt{\overline{{u'}^2}},\   v_{rms}=\sqrt{\overline{{v'}^2}},\   w_{rms}=\sqrt{\overline{{w'}^2}}.
\end{equation}
Comparison is made in term of two anisotropy measures $v_{rms}/u_{rms}$ and $v_{rms}/u_{rms}$. In case C2, FST has a somewhat higher level of anisotropy after leaving the forcing region, where after a certain adjustment length, they become closer to unity and more isotropic.\medskip


The integral length scale is defined as a measure of the largest separation distance over which components of the eddy velocities at two distinct points are correlated.  The longitudinal $\Lambda_x$ and transverse $\Lambda_z$ integral length scales, are here determined through direct integration of their corresponding correlation functions 
\[\Lambda_x= U_{\infty} \int_{0}^{\infty} u(t)u(t+\Delta t)/u_{rms}^2dt,\quad \Lambda_z=  \int_{0}^{\infty} u(z)u(z+\Delta z)/u_{rms}^2dz.\]
%
Taylor’s hypothesis of frozen turbulence is used to convert the time scales to longitudinal scale. In order to get the value of integral length scales, integrals are truncated, where we used the zero crossing of correlation functions as the truncation value \citep{Kurian_2009}.



\begin{figure}
    \centering
\subfloat[]{    \includegraphics[trim={0cm 0cm 0cm 0cm},clip,width=0.48\textwidth]{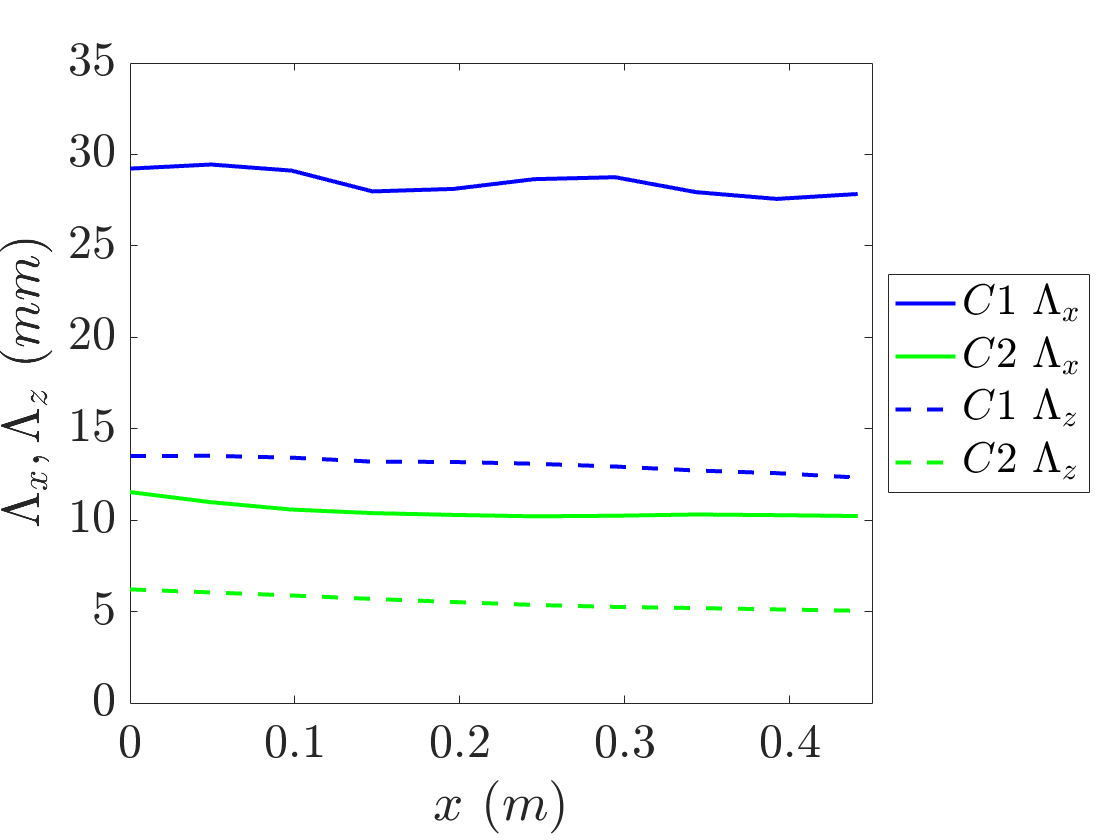}}
\subfloat[]{    \includegraphics[trim={0cm 0cm 0cm 0cm},clip,width=0.48\textwidth]{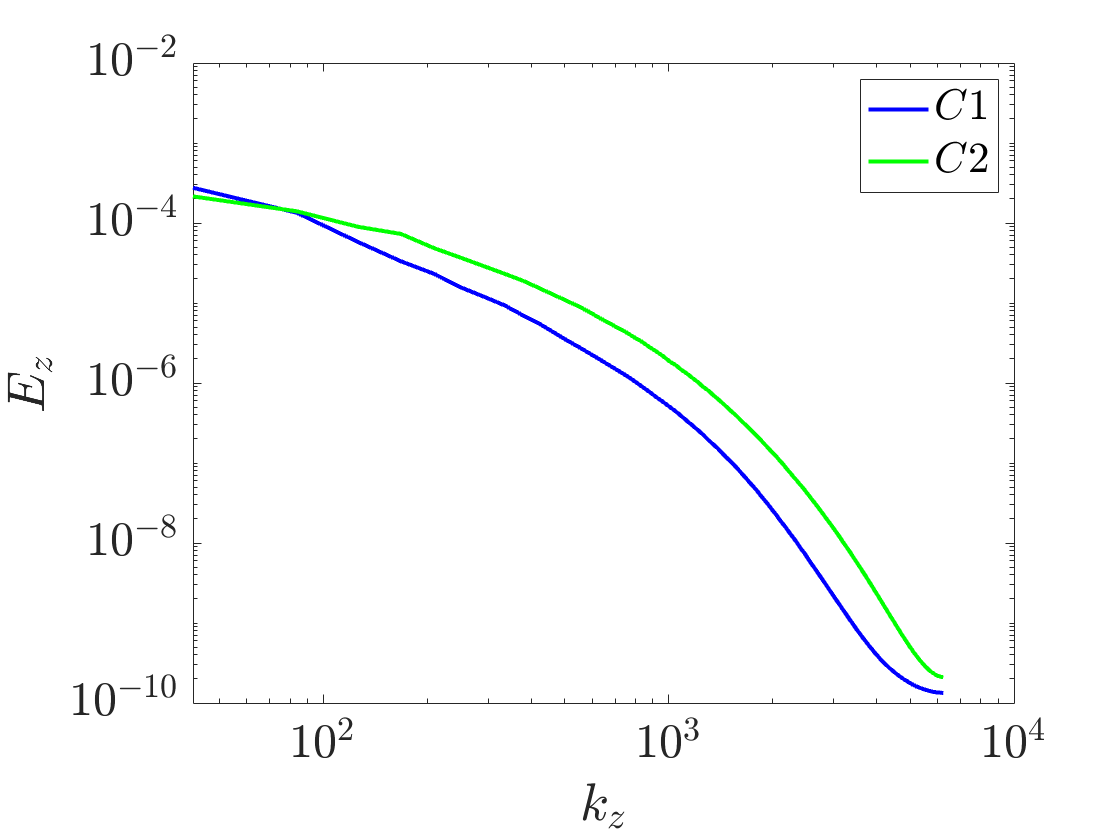}}
    \caption{(a) Integral length scale (streamwise $\Lambda_x$ and spanwise $\Lambda_z$) distribution along the streamwise direction. (b) One-dimensional energy spectrum in the spanwise direction in the free-stream at the location of the leading edge.}
    \label{fig:LSpectra}
\end{figure}

In figure \ref{fig:LSpectra}a streamwise and spanwise length scale development along the streamwise direction are shown. It can be seen that both streamwise and spanwise scales are keeping approximately constant values throughout the domain (with a small decay). Notice that the spanwise length scale is around half of the streamwise one, which is in accordance with isotropic turbulence assumption \citep{batchelor}. Calculated values of the length scales decrease from C1 to C2. Figure \ref{fig:LSpectra}b shows the comparison of the one-dimensional spectra in the free-stream at a streamwise position of the leading edge for both cases. Note that case with the larger integral length scale (C1) has larger energy for the very lowest spanwise wavenumbers, whereas the case with lower integral length scale has larger energy for intermediate and higher spanwise wavenumbers. This seemingly small difference is the main reason for the different transition lenghths in the two cases, which we will discuss at length in subsequent sections. 

\subsection{Mean flow}

In this section, the mean flow results for studied cases are compared and discussed. The statistics of the DNS data are based on time- and spanwise-averaging.

\begin{figure}
    \centering
\subfloat[]{       \includegraphics[width=0.48\textwidth]{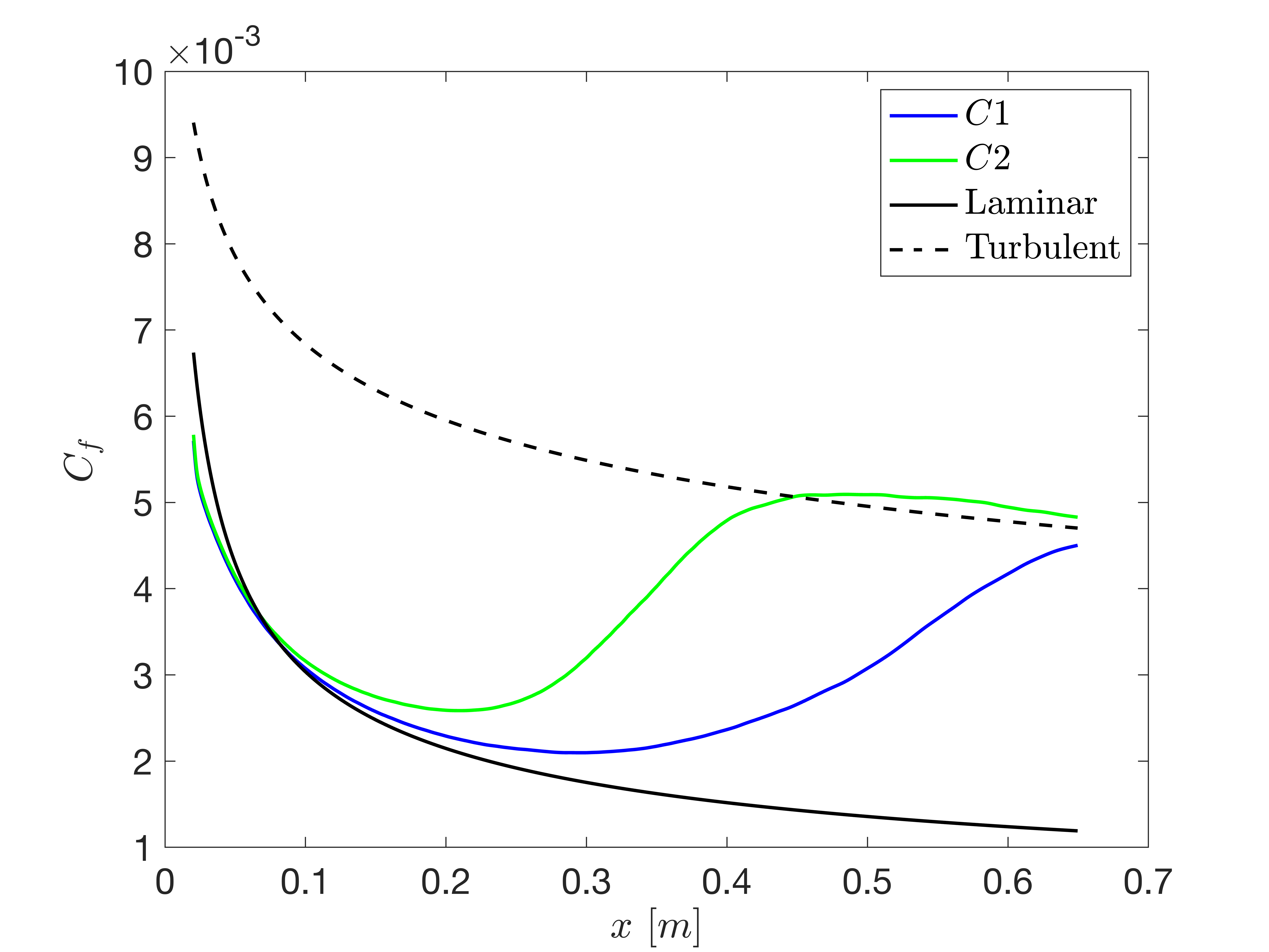}}
\subfloat[]{        \includegraphics[width=0.48\textwidth]{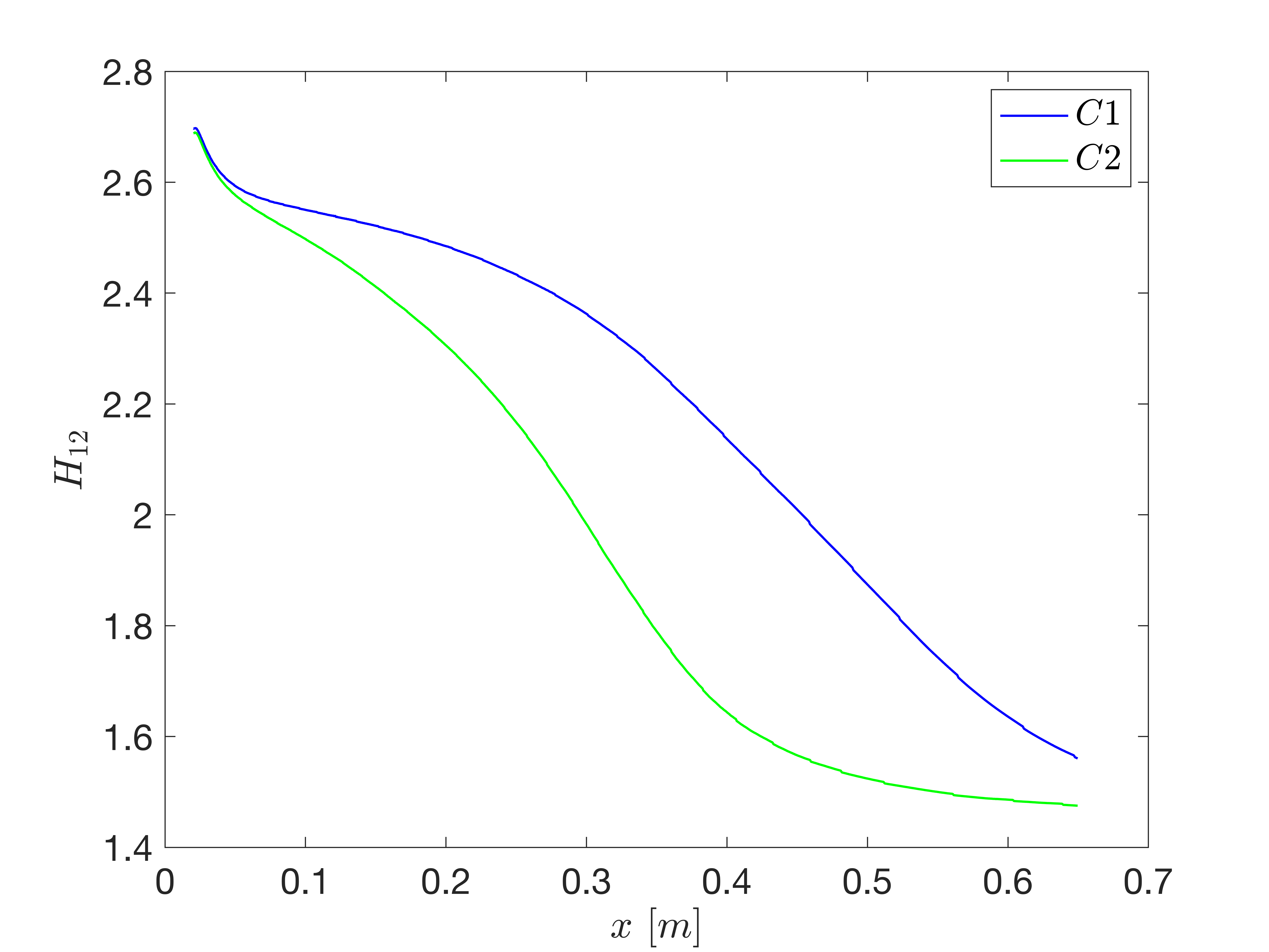}}
    \caption{(a) Skin friction distribution along the streamwise direction. Black lines represent laminar and turbulent flow given by $c_{f,lam}= 0.664 Re_x^{-1/2}$ and $c_{f,turb}=0.059 Re_x^{-1/5}$. (b) Downstream variation of the shape factor.}
    \label{fig:cfRe}
\end{figure}

The downstream development of the mean skin-friction coefficient, $c_f$ for different cases is presented in figure \ref{fig:cfRe}a.  Also shown are curves depicting laminar and turbulent flow given by $c_{f,lam}= 0.664 Re_x^{-1/2}$ and $c_{f,turb}=0.059 Re_x^{-1/5}$. Skin friction coefficient provides a good overview of the onset and length of transition. Despite a similar free-stream turbulence intensities, the two cases have different transition locations due to differences in their spectra and integral length scales at the inlet. We could define the onset of transition as the location where $c_f$ is at its minimum state, and completion of transition is where $c_f$ is maximum. The transition occurs earlier for the case with the lowest integral lengths scale. The flow remains transitional within the computational domain for the large integral length scale case, while the other case fully transition to a turbulent state. The skin friction is lower than the Blasius value for streamwise positions closer to the leading edge. This streamwise position corresponds to the leading-edge region with a slight adverse pressure gradient 
which causes the boundary layer to thicken, which lowers the skin friction. 

Figure \ref{fig:cfRe}b shows the spanwise-time-averaged shape factor, $H_{12}$ defined as the ratio of displacement thickness, $\delta ^*,$ and momentum thickness, $\theta$. The shape factor gives a good quantitative assessment of the mean streamwise velocity profile independent of the wall-normal derivatives. Moving from the leading edge downstream, initially, the shape factors in all simulations are identical and assume values of 2.6 that are typical for a laminar boundary layer. After this, the shape factors start to decrease as the boundary layer flow transitions to turbulence. The shape factor decreases to 1.4, characterising a fully turbulent state for case C2, while in case C1, the boundary layer is still in the transitional region at the end of the computational domain. 

\subsection{Boundary-layer perturbations}

\begin{figure}
    \centering
        \includegraphics[width=1.\textwidth]{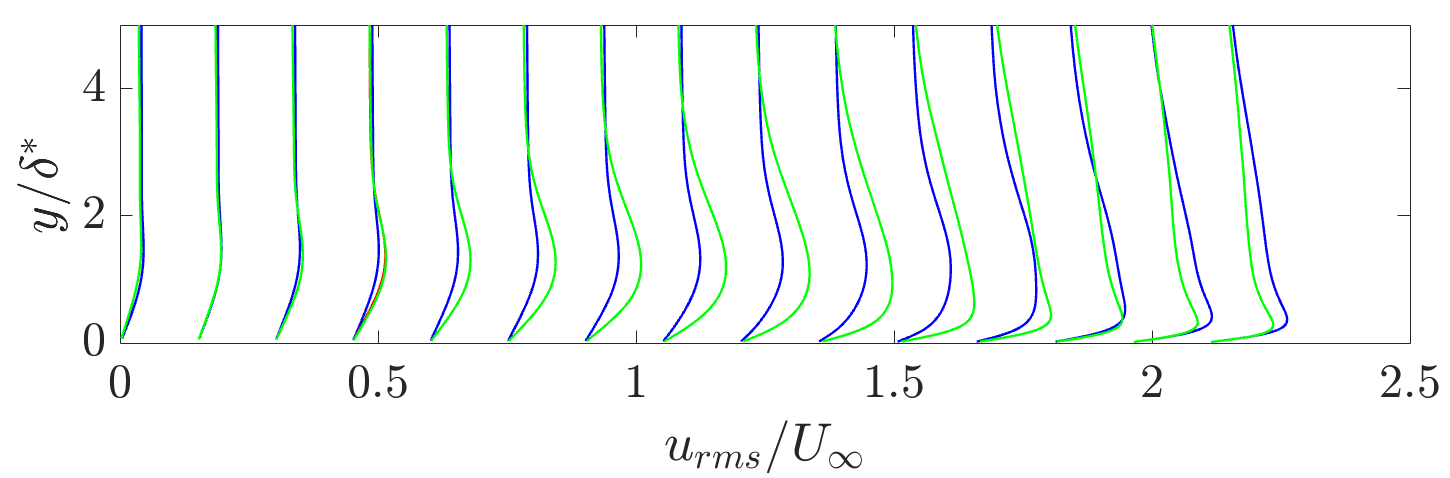}
    \caption{Root-mean-square streamwise velocity profiles shown at different streamwise locations, where $x$ starts at $0.01m$ and increases by $0.04m$. Streamwise, wall-normal and spanwise component are presented from top to bottom. The rms velocity profiles are shifted according to $u_{rms}/U_{\infty} = u_{rms}/U_{\infty} + b$ where $b$ increases by $0.15$.}
    \label{fig:rmsprof}
\end{figure}

In order to gain more insight into the nature of the boundary layer perturbations caused by FST, the second-order moment (root-mean-square values) were computed. These values, as a function of the wall-normal distance normalised by the local displacement thickness $\delta ^*$ for different downstream locations, are presented in figure \ref{fig:rmsprof}. Moving downstream, the peak values of $u_{rms}$ increase while location of the maximum value moves towards the wall.  The r.m.s.-profile scales approximately with $\delta ^*$, and its maximum is near $y/\delta ^* = 1.4$.

\begin{figure}
    \centering
\includegraphics[width=0.6\textwidth]{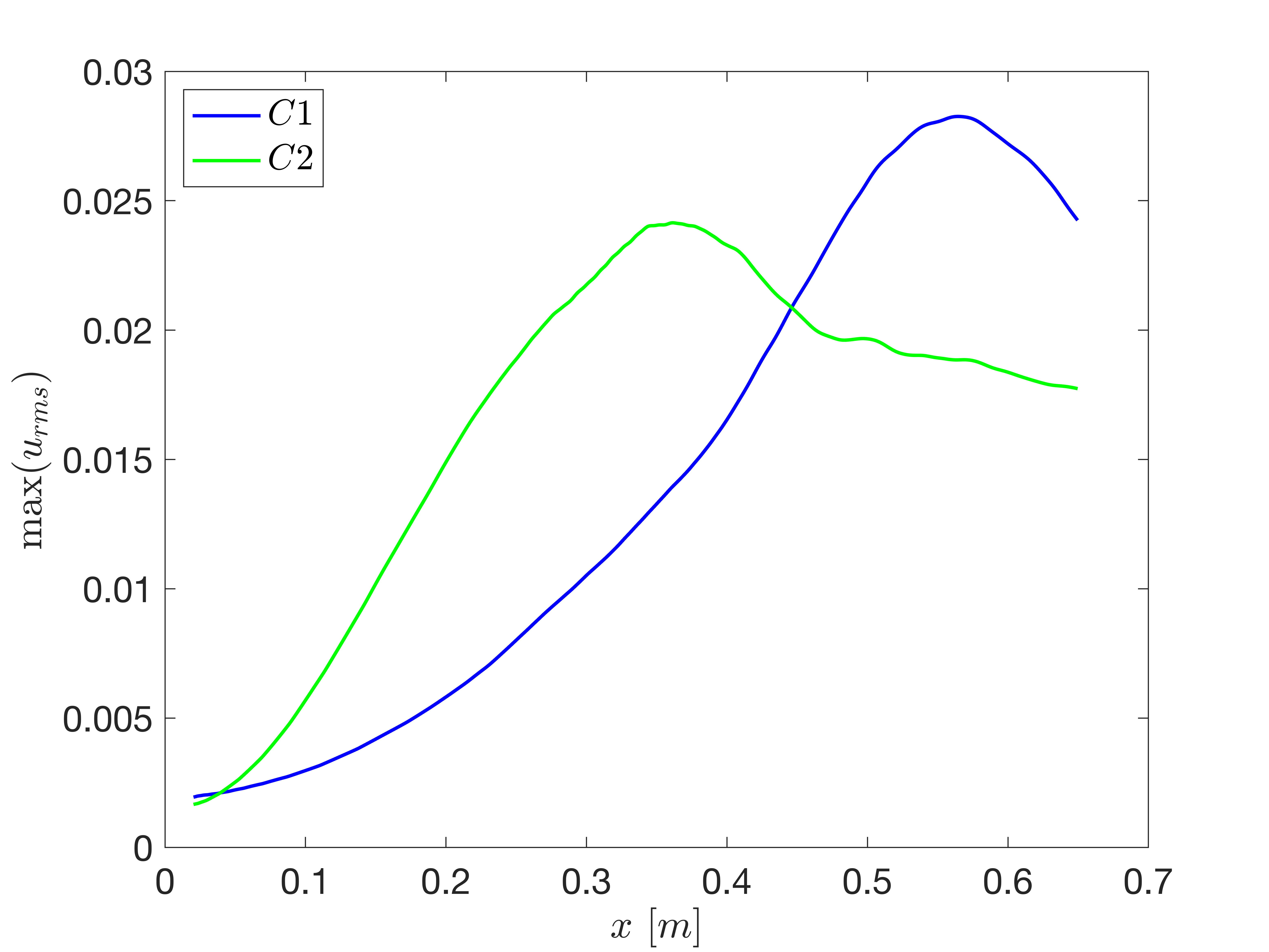}
    \caption{Maximum $u_{rms}$ of the perturbations inside the boundary layer. }
    \label{fig:Eh12}
\end{figure}

Maximum $u_{rms}$ of the perturbations inside the boundary layer is shown in figure \ref{fig:Eh12}. The general behavior seen there is that $u_{rms}$ grows in the downstream direction until it reaches a peak at some downstream location followed by a decay. The observed peak values move downstream as the integral length increases. This peak location seems to correspond to the skin friction curve halfway between the laminar and turbulent state (see figure \ref{fig:cfRe}a). For both cases, a rapid increase in energy is observed upstream of the transitional region, corresponding to the growing amplitude of streaks. Note that the case with lower energy in the largest spanwise wavnumbers increase much faster initially than the the case with more energy in the lowest wavenumbers. We will see that this will be an important factor in the in the faster transition to turbulence for case C2, compared to case C1.

\subsection{Flow structures}
  \begin{figure}
    \centering
        \includegraphics[trim={1cm 6.5cm 0cm 4cm},clip,width=\textwidth]{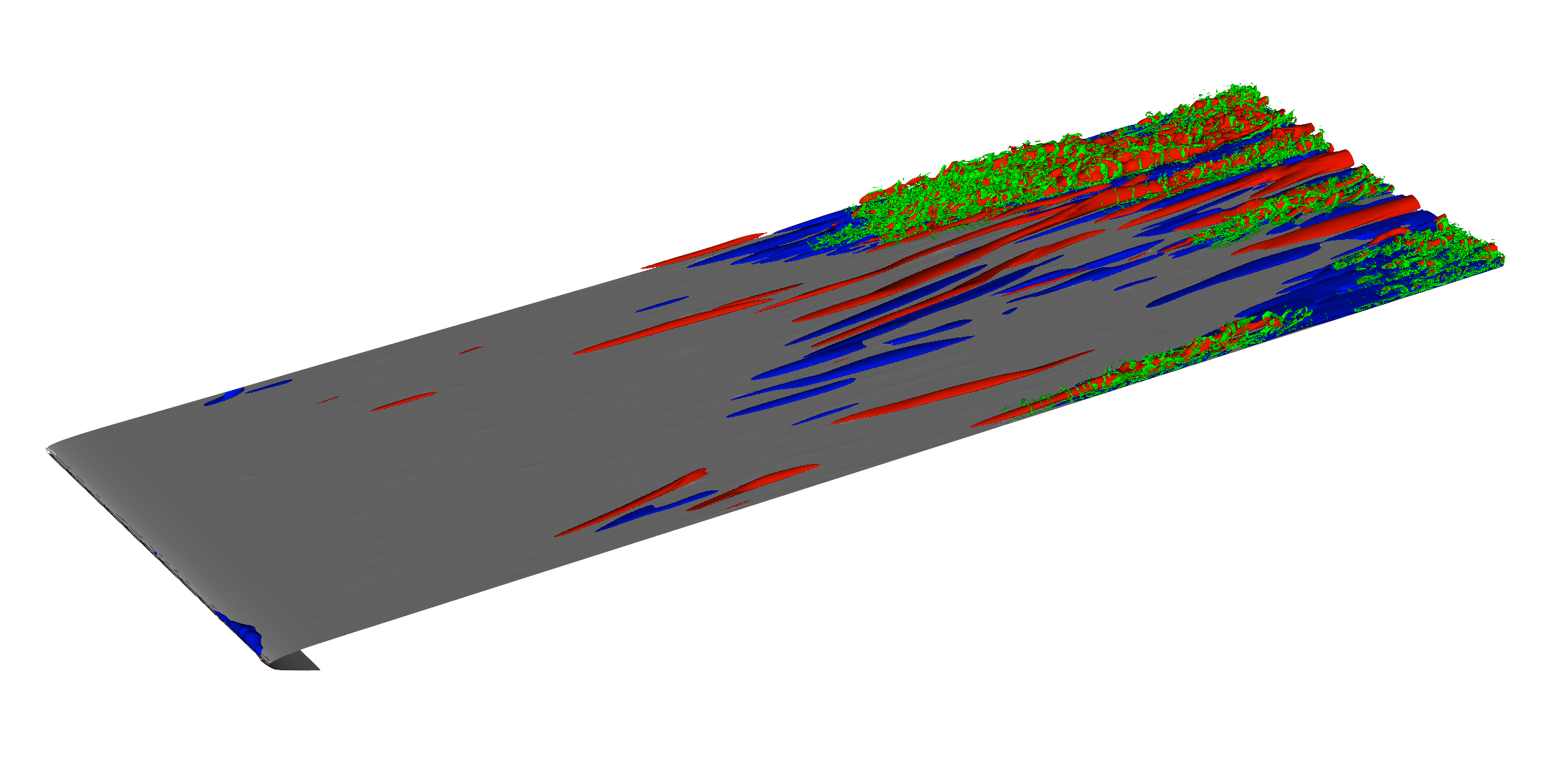}\\
        \vspace{-1.5cm}
        \includegraphics[trim={0cm 5cm 0cm 4cm},clip,width=\textwidth]{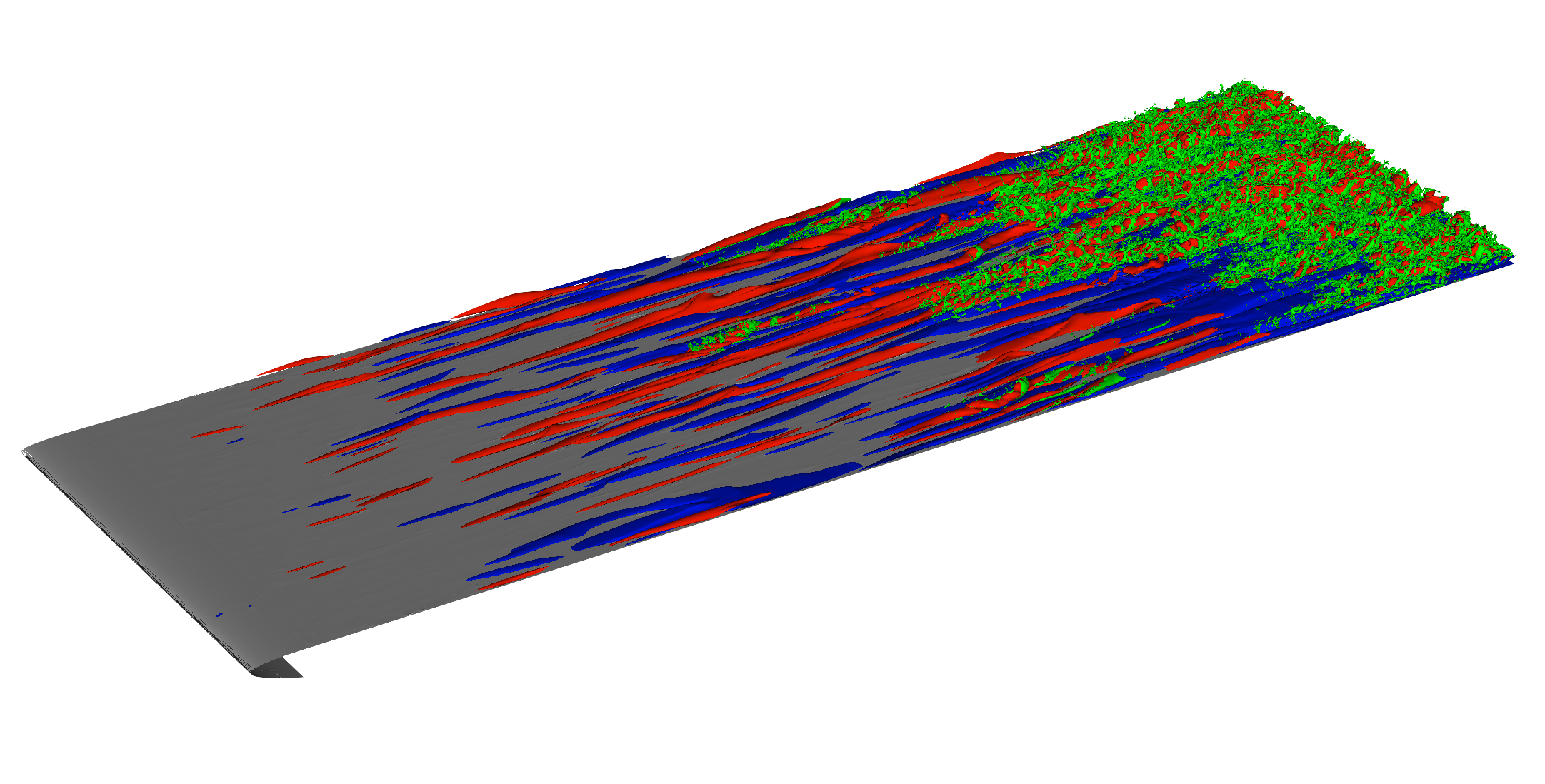}
    \caption{Snapshots of streamwise disturbance velocity component in the boundary layer. $\lambda_2$ structures are represented in green color, while red and blue mark the positive and negative disturbance velocity $u'=\pm 0.13U_{\infty}$, respectively. From top to bottom, C1, C2.}
    \label{fig:lambda2}
\end{figure}
A snapshot of the flow is shown in figure \ref{fig:lambda2} where the instantaneous streamwise disturbance velocities are presented with $\lambda_2$ structures \citep{lambda2}. Positive disturbance velocity is shown in red, negative in blue, while $\lambda_2$ structures are shown in green color. The overall picture of the transition scenario can be deduced from figure \ref{fig:lambda2}. Starting from the leading edge position, the perturbations inside the boundary layer appear mainly in the streamwise velocity component, in the form of elongated structures. Patches of irregular motion in the form of turbulent spots are seen to appear further downstream. As they travel downstream, the spots become wider and longer. The turbulent region at the end of the domain is created by the enlargement and merging of the various spots, and therefore the streamwise position at which the flow is turbulent varies with time. This difference between the cases clearly correspond to the difference in disturbance energy seen in figure \ref{fig:Eh12}.

\begin{figure}
    \centering
        \includegraphics[trim={3cm 0cm 55cm 1cm},clip,width=0.48\textwidth]{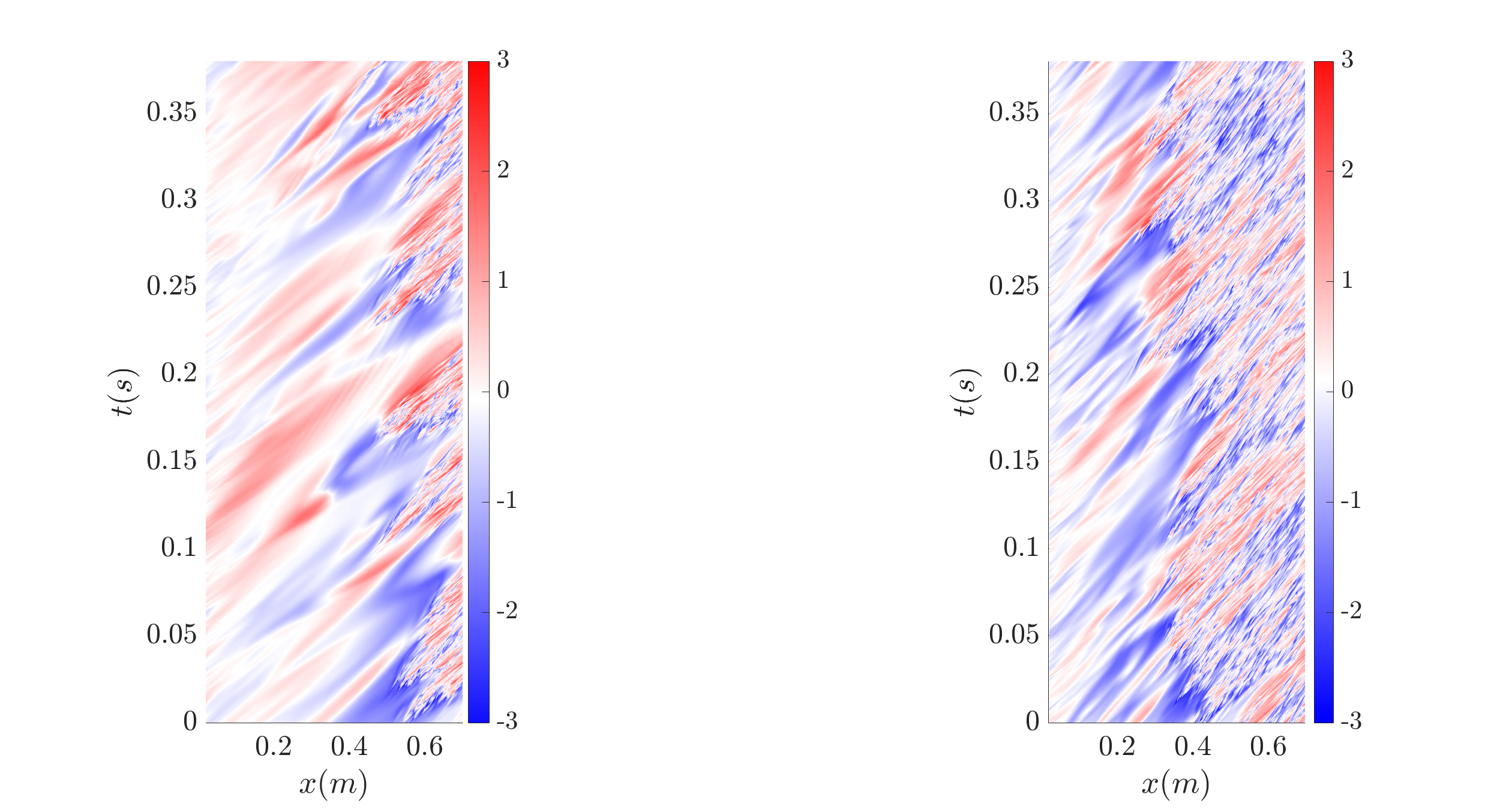}
        \includegraphics[trim={54cm 0cm 4cm 1cm},clip,width=0.48\textwidth]{img/timespace_allcases.png}
   \caption{Time-space diagrams showing disturbance velocity for C1, C2 (from left to right). Time history is shown for data points $z=0$ and wall-normal position of maximum $u_{rms}$.}
    \label{fig:timespace}
\end{figure}

Figure \ref{fig:timespace} shows the time-space variation of instantaneous, streamwise disturbance velocity. Results are shown for the spanwise location of $z=0\ m$ (middle of the spanwise domain), at the height of maximum $u_{rms}$. Red areas correspond to the positive disturbance velocity, and blue areas have negative disturbance velocity. We can observe the differences in streak sizes, as well as transition location for the studied cases. As also seen earlier, transition happens at  further downstream position in the case with the largest integral length scale (C1) compared to the one with the smallest integral length scale (C2). Note that it appears that the spanwise scale of the streaks decreases for case C1 as the transtion location is approached. 

\begin{figure}
    \centering
    \includegraphics[width=\linewidth]{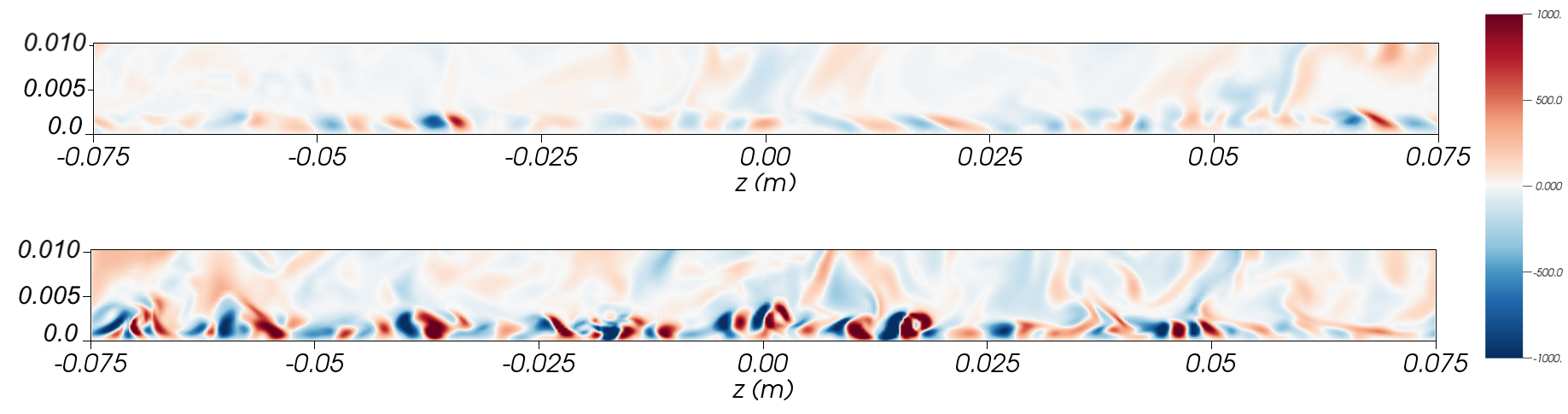}
    \caption{Wall-normal plane view of wall-normal vorticity at streamwise position $x=0.2m$, for C1 and C2 (from top to bottom). }
    \label{fig:vortiyz}
\end{figure}

This is more apparent in figure \ref{fig:vortiyz} which presents an alternative visualisation of the streamwise streaks. Here, the instantaneous wall-normal vorticity contours are plotted in the $z-y$-plane for cases C1 (top) and C2 (bottom). The plane is located in the pre-transitional region at $x = 0.2\ m$. For the case with the highest integral length scale (C1), a smaller number of streamwise streaks can be observed, however the vorticity contours of the strongest streaks appear to have similar spacing as for the more intense streaks seen in case C2. 

\begin{figure}
    \centering
        \includegraphics[trim={0cm 0cm 1cm 0cm},clip,width=0.48\textwidth]{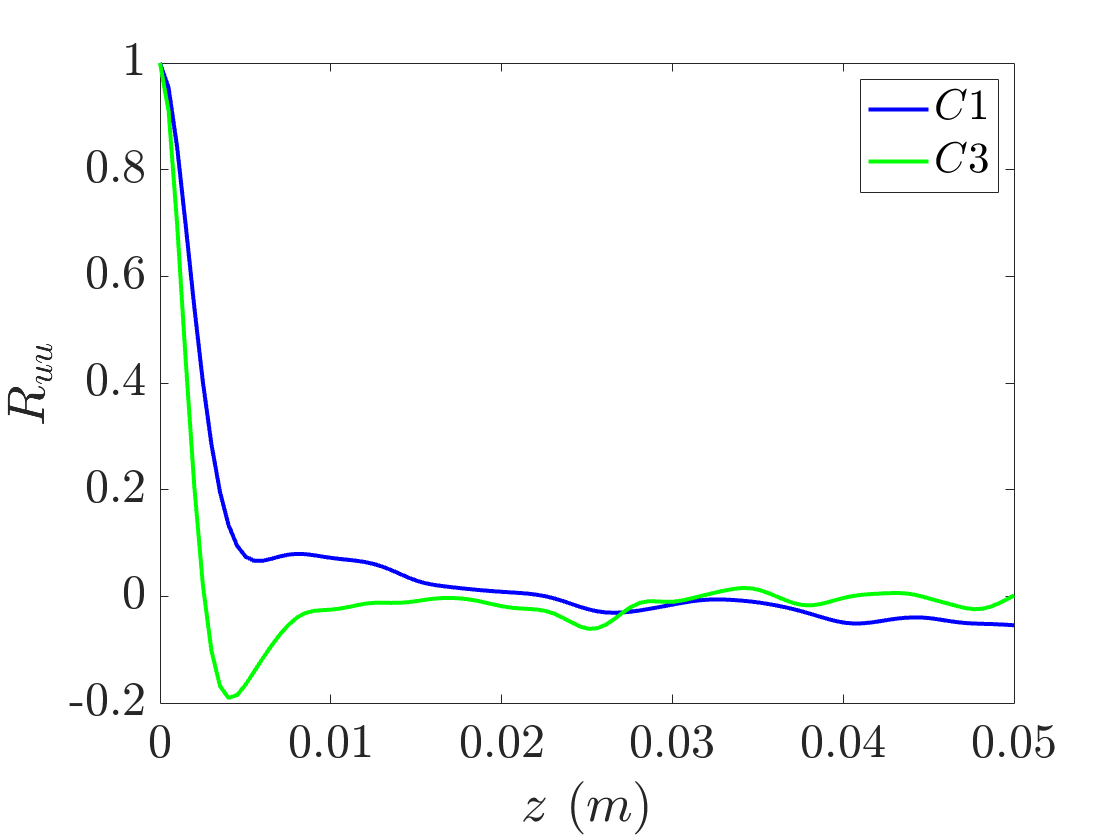}
        \includegraphics[trim={0cm 0cm 1cm 0cm},clip,width=0.48\textwidth]{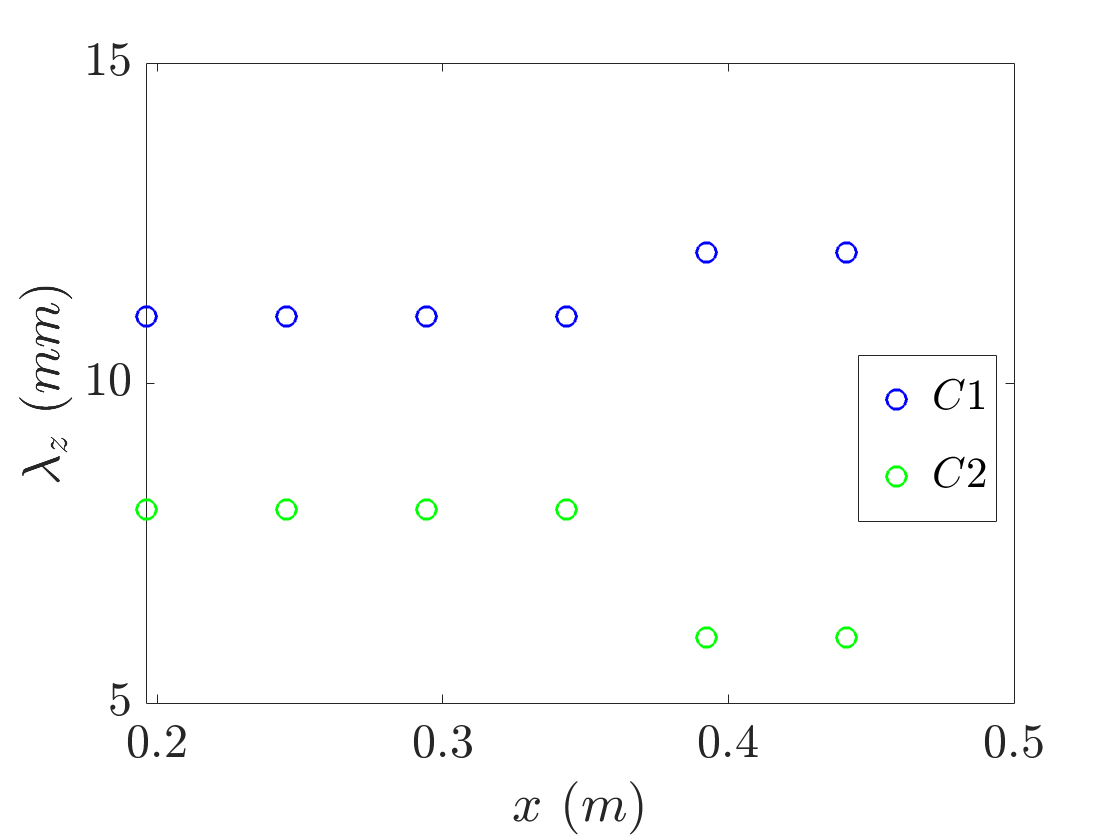}
    \caption{Two-point correlation in the boundary layer at $x=0.3m$  and calculated streak spacing as a function of streamwise distance. Measurements were taken at the wall-normal position of maximum $u_{rms}$.}
    \label{fig:streaksp}
\end{figure}

Streamwise elongated streaks emerge quite generally as a feature of zero-pressure-gradient laminar boundary layers under free-stream eddies. Figure \ref{fig:streaksp}(left) shows the spanwise correlation in the boundary layer. The averaged streak spacing is often defined as twice the distance to the first minimum of the correlation function. In figure \ref{fig:streaksp}(right), we show the comparison of obtained values for average stream spacing for all cases. These results are consistent with experimental measurements of \cite{fransson_shahinfar_2020}, and this type of spanwise structure is absent in the free stream. Note that the 
streak spacing is increasing slightly with increasing value of the free-stream turbulent integral length scale, but are surprisingly similar in spanwise width considering the difference in the initial scales closer to the leading edge. This similarity in scales can also be seen in figure \ref{fig:lambda2} for the instantaneous streamwise disturbance velocities.

\section{Influence of large integral length scales}
\label{Influence}

In order to better understand the differences between scales of the boundary-layer structures generated by the free-stream disturbances in cases C1 and C2, we again consider figure~\ref{fig:LSpectra}. 



The freestream turbulence close to the leading edge in case C1 is concentrated at lower spanwise wavenumbers (larger spanwise structures) compared to those in case C2, where the energy is spread more towards higher spanwise wavemubers. This data is consistent with the time history presented in figure~\ref{fig:timespace}, where there are clearly structures with larger spanwise scales close to the leading edge for case C1 compared to C2. This means that the receptivity process, where the freestream purturbations enter into the boundary layer, as expected, do not change the spanwise distribution of the spectral energy to any large extent.


%

\begin{figure}
    \centering
    \includegraphics[width=0.4\textwidth]{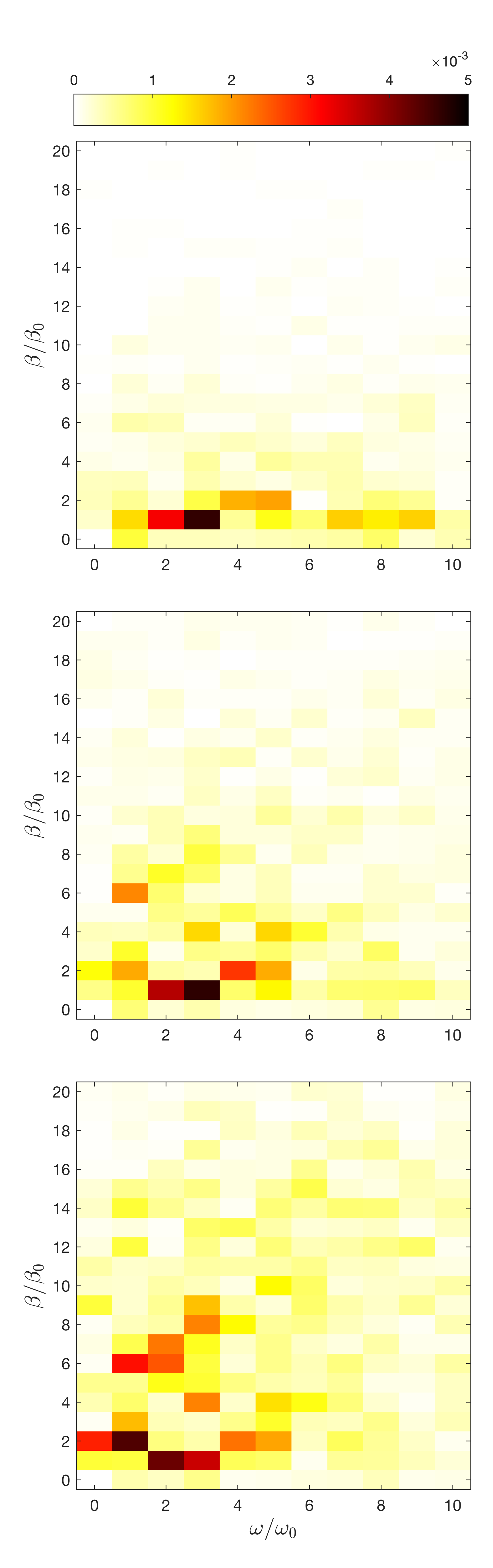}
    \includegraphics[width=0.4\textwidth]{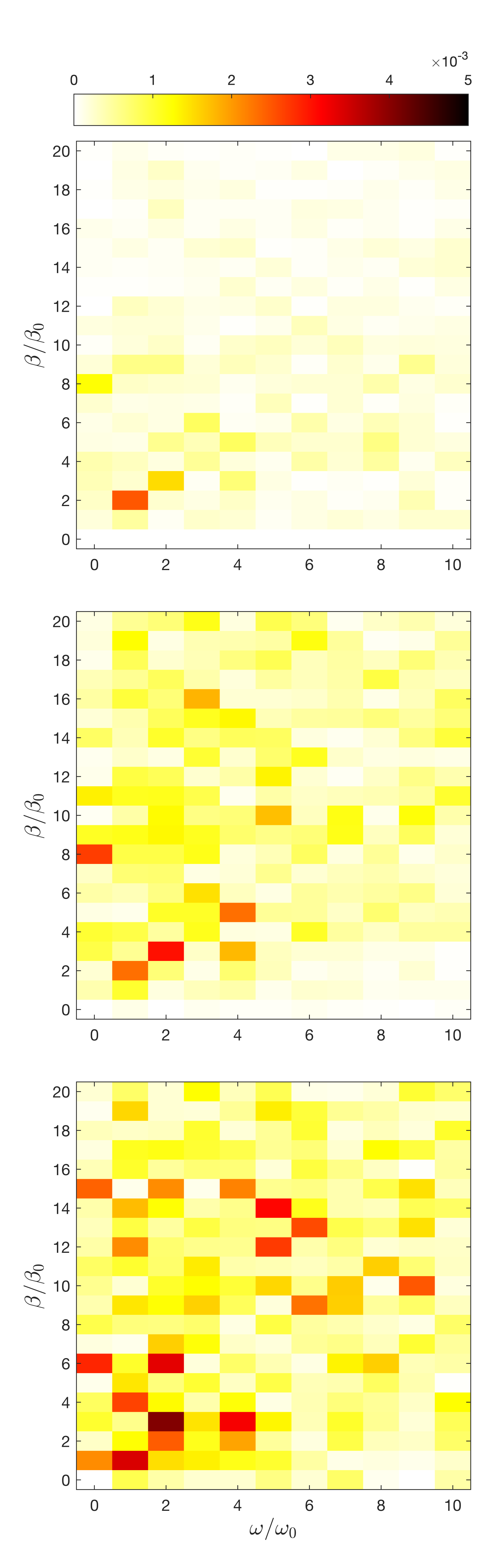}    
    \caption{Energy of the different Fourier components of disturbance filed, inside the boundary layer (top to bottom at $x=0.1,\ 0.2 $ and $0.3\ m$. Left column corresponds to case C1 ($\Lambda_x=29.22\ mm$) and right one to case C2 ($\Lambda_x=11.53\ mm$). $\omega_0=12.76\ rad/s$ for case C1 and $12.19\ rad/s$ for case 2, $\beta_0=41.89\ m^{-1}$ for both cases.}
    \label{fig:fft_x15}
\end{figure}

\begin{figure}
    \centering
    \includegraphics[width=0.6\textwidth]{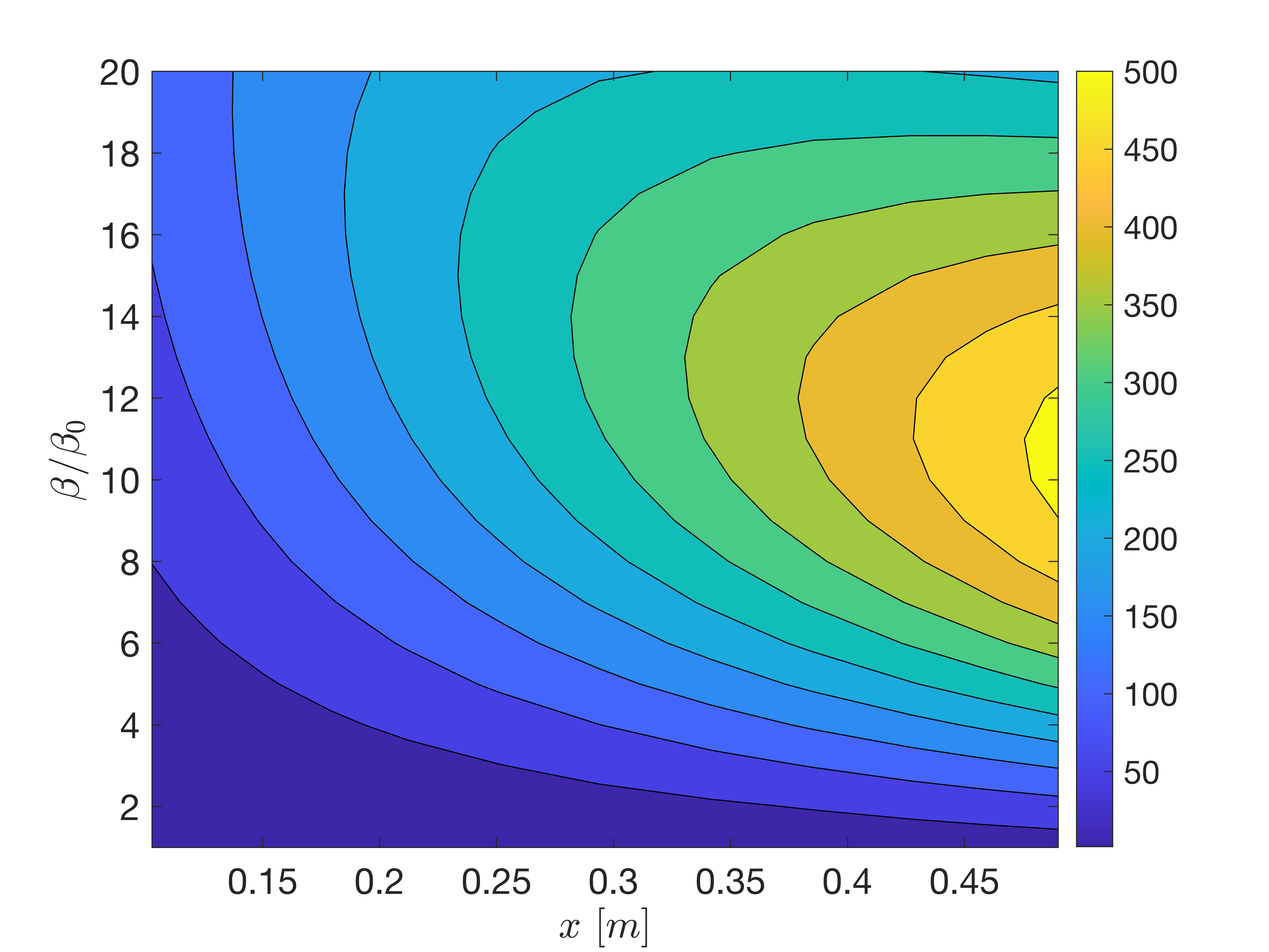}
\caption{Optimal energy growth as a function of streamwise position and spanwise wavenumber for steady perturbations.}
    \label{fig:opt_growth}
\end{figure}

To understand the disturbance development of the perturbations for the two cases in more detail, figure \ref{fig:fft_x15} shows the two-dimensional energy spectra of the different Fourier components of the disturbances at three distances close to the leading edge. In the left three figures, corresponding to case C1, we see that the energy is located closer to the lowest spanwise wavenumbers and propagates less to higher wavenumbers, than that for the case C2 seen in the right three figures. The figures are consistent with the $u_{rms}$ growth seen in figure \ref{fig:Eh12} and the appearance of the flow structures in figure \ref{fig:lambda2}. In the right lower figure \ref{fig:fft_x15} we can see that a large part of the energy has reached the spanwise wavenumber $15\beta_0$, which corresponds to the same spanwise scale seen in the streaks seen in figure \ref{fig:vortiyz}(bottom). Those streaks are the ones that subsequently break down to turbulence further downstream. For the lower left figure \ref{fig:fft_x15} we do not see any substantial energy in similar wavenumbers, but can expect that this will occur further downstream based on the fact that similar spanwise scales are seen in the streaks in figure \ref{fig:lambda2}(top).

The differences in the streamwise development in the two cases can be understood if we consider the largest growth possible for streaky disturbances in the boundary layers of laminar flow seen in figure \ref{fig:opt_growth}, plotted for the parameters of these simulations. Here, the energy gain 
\begin{equation}
G(x_f)=\max_\beta \frac{\int \mathbf{u}_f'^H\mathbf{u}_f'\ dy}{\int \mathbf{u}_0'^H\mathbf{u}_0'\ dy},
\end{equation}
for stationary ($\omega=0$) optimal perturbations initiated at $x_0=0.005\ m$ is computed as a function of final position $x_f$, following the works by \cite{andersson1999optimal} and \cite{Luchini}.

 Although the disturbances present close to the leading edge will not be identical to the optimal perturbations calculated here, the maximum growth calculated gives a very good indication of the growth possibilities for the various spanwise wavenumbers. How much of the optimal growth is achieved for the streaks in the simulations depends on how large the projection of the optimal disturbances are on the flow close to the leading edge. \cite{FaundezJFM2022} made such a calculation at the leading edge of a NACA0008 wing subjected to freestream turbulence, and found that there is a substantial projection on the optimal disturbance, thus showing that the maximum growth gives a good indication of what disturbances can be expected to grow in the boundary layer. As was seen above, for disturbances entering at the leading edge, nonlinearity is transferring energy to higher spanwise wavenumbers as the disturbances propagate downstream. As the energy enter regions of wavenumber space with larger optimal growth, they have a potential of utilizing the growth mechanisms at those wavemumbers.

We see that the growth initially is smaller for the largest integral length scale, since the major energy for that case is located at the very bottom of the figure, corresponding to a normalized spanwise wavenumber of one. The smaller integral length scale has its major energy close to the leading edge in one to two wave numbers larger, i.e. $\beta/\beta_0=2$ and 3. If the flow development only was governed by linear mechanisms these streaks would remain the only structures with major energy in the flow. However, the energy in the low spanwise wavenumbers are large enough for disturbance energy to rapidly propagate to smaller spanwise scales. Figure \ref{fig:opt_growth} shows that the maximum possible energy growth occur for spanwise wavenumbers around $15\beta_0$ for the streamwise positions around $0.2-0.3\ m$. This implies that should energy be able to propagate to those scales, it would quickly amplify in the boundary layer. Considering the streaks in figure \ref{fig:vortiyz}(bottom), we see that they have precisely that scale. The figures \ref{fig:fft_x15} to the right shows that the energy is in fact propagating to the correct spanwise wavenumbers to be able to utilize this growth possibility. For the case of larger spanwise scales the propagation of energy to higher spanwise wavenumbers takes a longer time and the region of maximal energy growth is reached later, implying that the larger amplitude streaks that subsequently break down appear further downstream. Note, however, that the spanwise scales of the streaks that break down, compared to the size of boundary layer, are similar for the two cases, corresponding in both cases to spanwise wave lengths where the maximum growth of the streaks occur for those streamwise positions. 

Thus, the spectral energy in the freestream, its propagation into the boundary layer through the receptivity process, and the subsequent growth possibilities of these streaks, in particular those non-linearly generated, can explain that the transition occurs later for larger integral length scales.

\section{Summary and conclusions}\label{sec:conclusion}
In the present study, we have performed DNS of the FST-induced boundary-layer transition on a flat plate with a realistic leading edge to investigate effects of large integral length scales of FST on boundary-layer transition. 

While the turbulence intensity is known to correlate with the transition Reynolds number strongly, characteristic length scales of the FST are often considered to have rather a low impact on the transition location. However, a recent experiment by \cite{fransson_shahinfar_2020} shows significant effects of FST scales. They found that for low turbulence intensity values, an increase of the integral length scale advances transition, which agrees with the literature. However, an increase in the integral length scale delays transition for higher turbulence intensities. Here, we have examined this finding for intermediate levels of freestream turbulence, corresponding to cases where increasing the integral length scale delays transition to turbulence. Our simulations supports this trend observed by \cite{fransson_shahinfar_2020}, and gives a flow physics explanation of the mechanisms behind this result. Contrary to many other DNS simulations, the ones presented here have a substantially larger integral length scale. This is important since we have shown that the flow structures with the largest spanwise length scales have to be well beyond the peak of maximum transient growth for this delay of transition with increasing length scale to occur. For such structures the energy needs to propagate to the smaller spanwise wavenumbers where the maximum growth occurs, and where the resulting large amplitude streaks are the ones that break down to turbulent spots. 

Comparing further to the results of \cite{fransson_shahinfar_2020} we cannot confirm their optimal ratio between the freestream scales and the aspect ratio of the streaks at the transition location. Rather we find that at the transition location for our flow cases the streak spacing are close to the wavemnumber for the optimal growth, implying that we agree more with what researchers previously have found. However, we have presented two different simulation cases and not such an extensive study as the one by \cite{fransson_shahinfar_2020}. Further high-fidelity simulations may be needed to understand this fully, since it is not always straight forward to determine the width of the streaks at transition experimentally. 

\section*{Acknowledgements}
We thank Jens Fransson for fruitful discussions and kindly providing data from the experiments. 
Financial support for this work was provided by the European Research Council under grant agreement 694452-TRANSEP-ERC-2015-AdG. The computations were performed on resources by the Swedish National Infrastructure for Computing (SNIC) at the PDC Center for High Performance Computing at the Royal Institute of Technology (KTH) and at National Supercomputer Centre at Link\"{o}ping University. We acknowledge PRACE for awarding us access to MareNostrum at Barcelona Supercomputing Center (BSC), Spain. 



\bibliography{biblio}

\end{document}